\documentclass[a4paper,11pt]{article}
\usepackage{pos}
\usepackage{slashed}
\usepackage{tikz-feynman}
\tikzfeynmanset{compat=1.0.0}

\newcommand{\si}[1]{s(#1)}
\newcommand{\co}[1]{c(#1)}
\newcommand{\sbar}[1]{\bar{s}(#1)}
\newcommand{\cbar}[1]{\bar{c}(#1)}

\newcommand{\csw}{{c_\text{SW}}}

\title{$\mathbf{c_\textbf{SW}}$ at One-Loop Order for Brillouin Fermions}
 \ShortTitle{$\csw$ at One-Loop Order for Brillouin Fermions}

\author*[a]{Maximilian Ammer}
\author[a,b]{Stephan Dürr}

\affiliation[a]{Physics Department, University of Wuppertal, D-42119 Wuppertal, Germany}

\affiliation[b]{IAS/JSC, Forschungszentrum Jülich, D-52425 Jülich, Germany}

\emailAdd{ammer(AT)uni-wuppertal.de}

\abstract{Wilson-like Dirac operators can be written in the form $D=\gamma_\mu\nabla_\mu-\frac {ar}{2} \Delta$. For Wilson fermions the 
standard two-point derivative $\nabla_\mu^{(\mathrm{std})}$ and 9-point Laplacian $\Delta^{(\mathrm{std})}$ are used. For 
Brillouin fermions these are replaced by improved discretizations $\nabla_\mu^{(\mathrm{iso})}$ and $\Delta^{(\mathrm{bri})}$ 
which have 54- and 81-point stencils respectively. We derive the Feynman rules in lattice perturbation theory for the Brillouin 
action and apply them to the calculation of the improvement coefficient $\csw$, which, similar to the Wilson case, has a 
perturbative expansion of the form $\csw=1+\csw^{(1)}g_0^2+\mathcal{O}(g_0^4)$. For $N_c=3$
we find ${\csw}^{(1)}_\text{Brillouin} =0.12362580(1) $, compared to ${\csw}^{(1)}_\text{Wilson} = 0.26858825(1)$, both for $r=1$.}

\FullConference{%
The 39th International Symposium on Lattice Field Theory,\\
8th-13th August, 2022,\\
Rheinische Friedrich-Wilhelms-Universität Bonn, Bonn, Germany
}

\begin{document}
\maketitle

\section{Introduction} 
The massless Wilson Dirac Operator can be expressed in terms of the standard two-point derivative $\nabla^\mathrm{std}_\mu$ and the standard 9-point Laplacian $\Delta^\mathrm{std}$
\begin{align}
D_W(x,y)=\sum\limits_\mu \gamma_\mu \nabla^\mathrm{std}_\mu(x,y)-\frac {ar}{2} \Delta^\mathrm{std}(x,y).
\end{align}
The massless Brillouin Dirac operator is obtained by replacing the derivative and Laplacian by different, more complicated discretizations called $\nabla^\mathrm{iso}_\mu$ and $\Delta^\mathrm{bri}$ \cite{Durr:2010ch}\cite{Durr:2021iff}
\begin{align}
D_B(x,y)=\sum\limits_\mu \gamma_\mu \nabla^\mathrm{iso}_\mu(x,y)-\frac {ar}{2} \Delta^\mathrm{bri}(x,y).
\end{align}
This way the Brillouin Dirac operator has an 81-point stencil containing all \emph{off-axis} points that are 1-,2-,3-, and 4-hops away from $x$. The contributing points are weighted by the coefficients
\begin{align}
(\rho_1,\rho_2,\rho_3,\rho_4)&=\frac{1}{432}(64,16,4,1)\\
(\lambda_0,\lambda_1,\lambda_2,\lambda_3,\lambda_4)&=\frac{r}{64}(-240,8,4,2,1),
\end{align}
such that
\begin{align}
D_B(x,y)=&-\frac{\lambda_0}{2}\delta(x,y)\nonumber\\
&+\sum\limits_{\mu=\pm1}^{\pm 4}\left(\rho_1\gamma_\mu-\frac{\lambda_1}{2}\right)W_\mu(x)\delta(x+\hat{\mu},y)\nonumber\\
&+\sum\limits_{\substack{\mu,\nu=\pm1\\ |\mu|\neq|\nu|}}^{\pm 4}\left(\rho_2\gamma_\mu-\frac{\lambda_2}{4}\right)W_{\mu\nu}(x)\delta(x+\hat{\mu}+\hat{\nu},y)\nonumber\\
&+\sum\limits_{\substack{\mu,\nu,\rho=\pm1 \\ |\mu|\neq|\nu|\neq|\rho|}}^{\pm 4}\left(\frac{\rho_3}{2}\gamma_\mu-\frac{\lambda_3}{12}\right)W_{\mu\nu\rho}(x)\delta(x+\hat{\mu}+\hat{\nu}+\hat{\rho},y)\nonumber\\
&+\sum\limits_{\substack{\mu,\nu,\rho,\sigma=\pm1 \\ |\mu|\neq|\nu|\neq|\rho|\neq|\sigma|}}^{\pm 4}\left(\frac{\rho_4}{6}\gamma_\mu-\frac{\lambda_4}{48}\right)W_{\mu\nu\rho\sigma}(x)\delta(x+\hat{\mu}+\hat{\nu}+\hat{\rho}+\hat{\sigma},y),
\end{align}
where $|\mu|\neq|\nu|\neq\hdots$ is used in a transitive way i.e. the sums are over indices with pairwise different absolute values.
The $W$s are the average of the products of the link variables  $U$ along the paths:
\begin{align}
W_\mu(x)&=U_\mu(x)\\
W_{\mu\nu}(x)&=\frac 12\left(U_\mu(x)U_\nu(x+\hat{\mu})+\mathrm{perm}\right)\\
W_{\mu\nu\rho}(x)&=\frac 16\left(U_\mu(x)U_\nu(x+\hat{\mu})U_\rho(x+\hat{\mu}+\hat{\nu})+\mathrm{perms}\right)\\
W_{\mu\nu\rho\sigma}(x)&=\frac 1{24}\left(U_\mu(x)U_\nu(x+\hat{\mu})U_\rho(x+\hat{\mu}+\hat{\nu})U_\sigma(x+\hat{\mu}+\hat{\nu}+\hat{\rho})+\mathrm{perms}\right).
\end{align}

\section{Feynman Rules of the Brillouin Action}
We have derived the Feynman rules of the Brillouin action using a computer algebra system (Mathematica), supplemented by some analytical calculations by hand.

\subsection{The Fermion Propagator}
The Fourier transforms of the "free" derivative $\nabla_\mu^\mathrm{iso}$ and Laplace operator $\Delta^\mathrm{bri}$ are \cite{Durr:2010ch}:
\begin{align}
\nabla_\mu^\mathrm{iso}(k)&=\frac i{27} \sin(k_\mu)\prod\limits_{\nu\neq \mu}(\cos(k_\nu)+2)\\
\Delta^\mathrm{bri}(k)&=4\left(\cos^2(\tfrac 12 k_1)\cos^2(\tfrac 12 k_2)\cos^2(\tfrac 12 k_3)\cos^2(\tfrac 12 k_4)-1\right),
\end{align}
where $k_\mu$ is the fermion momentum in lattice units.
Then the Fermion propagator is
\begin{align}
&S_\mathrm{Brillouin}(k)=a\left(\sum\limits_\mu \gamma_\mu \nabla_\mu^\mathrm{iso}(k)-\frac {r}2 \Delta^\mathrm{bri}(k)\right)^{-1}
=a\frac{-\sum_\mu \gamma_\mu \nabla_\mu^\mathrm{iso}(k)-\frac {r}{2} \Delta^\mathrm{bri}(k)}{\frac {r^2}4 \Delta^\mathrm{bri}(k)^2-\left(\sum_\mu  \nabla_\mu^\mathrm{iso}(k)^2\right)}.
\end{align}
\subsection{The Vertices}
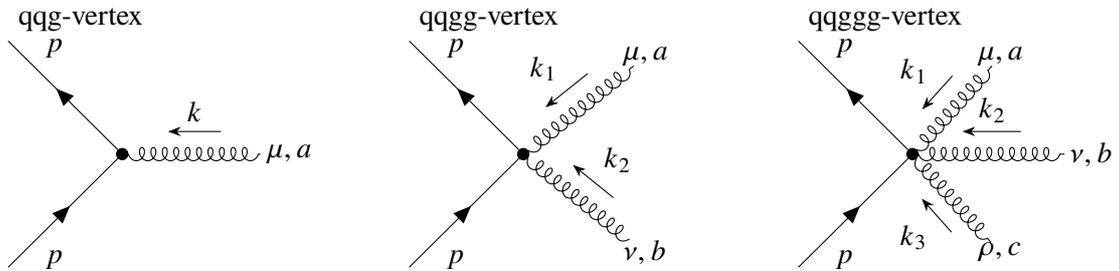
\begin{figure}[!b!]
\centering
\begin{tikzpicture}
  \begin{feynman}
  \vertex[draw,circle,fill,inner sep=1.5pt] (a){};
  \vertex [right= 2.2cm of a, inner sep=2pt] (b){$\mu,a$};
  \vertex [above left = 1.5cm and 1.5cm of a] (i1);
  \vertex [below left = 1.5cm and 1.5cm of a] (i2);
  \diagram*{
   (i2) --[fermion, edge label'=$p$, near start] (a),
   (a) -- [fermion, edge label'=$p$, near end] (i1),
   (a) -- [gluon,rmomentum={[arrow shorten=0.3]$k$}] (b),
   };
   \node[above,anchor=south west] at (i1) {qqg-vertex};
  \end{feynman}
\end{tikzpicture}
\hspace{1cm}
\begin{tikzpicture}
  \begin{feynman}
  \vertex[draw,circle,fill,inner sep=1.5pt] (a){};
  \vertex [above right= 1.3cm and 1.3cm of a, inner sep=0.cm,right] (b){$\mu,a$};
   \vertex [below right= 1.3cm and 1.3cm of a, inner sep=0.cm, right] (c){$\nu,b$};
  \vertex [above left = 1.5cm and 1.5cm of a] (i1);
  \vertex [below left = 1.5cm and 1.5cm of a] (i2);
  \diagram*{
   (i2) --[fermion, edge label'=$p$, near start] (a),
   (a) -- [fermion, edge label'=$p$, near end] (i1),
   (a) -- [gluon,rmomentum={[arrow shorten=0.3]$k_1$}] (b),
   (a) -- [gluon,rmomentum={[arrow shorten=0.3]$k_2$}] (c),
   };
   \node[above,anchor=south west] at (i1) {qqgg-vertex};
  \end{feynman}
\end{tikzpicture}
\hspace{1.5cm}
\begin{tikzpicture}
  \begin{feynman}
  \vertex[draw,circle,fill,inner sep=1.5pt] (a){};
  \vertex [right= 2.cm of a, right,inner sep=1.5pt] (b){$\nu,b$};
  \vertex [above right= 1.3cm and .85cm of a, inner sep=0.cm,right] (c){$\mu,a$};
   \vertex [below right= 1.3cm and .85cm of a, inner sep=0.cm, right] (d){$\rho,c$};
  \vertex [above left = 1.5cm and 1.5cm of a] (i1);
  \vertex [below left = 1.5cm and 1.5cm of a] (i2);
  \diagram*{
   (i2) --[fermion, edge label'=$p$, near start] (a),
   (a) -- [fermion, edge label'=$p$, near end] (i1),
   (a) -- [gluon,rmomentum={[arrow shorten=0.3]$k_2$}] (b),
   (a) -- [gluon,rmomentum={[arrow shorten=0.3]$k_1$}] (c),
   (a) -- [gluon,rmomentum'={[arrow shorten=0.3]$k_3$}] (d),

   };
   \node[above,anchor=south west] at (i1) {qqggg-vertex};
  \end{feynman}
\end{tikzpicture}
\caption{Momentum assignments for the vertices with one, two and three gluons. }\label{figvertices}
\end{figure}

For the calculation of $\csw$ to one-loop order below the three vertices with one, two and three gluons coupling to a quark and anti-quark are needed (see Figure \ref{figvertices}).
We define the following notations to express the vertex Feynman rules:
\begin{align}
\si{k_\mu}&=\sin\left(\tfrac 12 k_\mu\right)& \co{k_\mu}&=\cos\left(\tfrac 12 k_\mu\right)\\
\sbar{k_\mu}&=\sin\left( k_\mu\right)& \cbar{k_\mu}&=\cos\left( k_\mu\right)
\end{align}
\begin{align}
K^{(fg)}_{\mu\nu}(p,q)&=f(p_\mu+q_\mu)\left[\bar{g}(p_\nu)+\bar{g}(q_\nu)\right]\\
K^{(fgh)}_{\mu\nu\rho}(p,q)&=f(p_\mu+q_\mu)\left\{\bar{g}(p_\nu)\bar{h}(p_\rho)+\bar{g}(q_\nu)\bar{h}(q_\rho)+[\bar{g}(p_\nu)+\bar{g}(q_\nu)][\bar{h}(p_\rho)+\bar{h}(q_\rho)]\right\}\\
K^{(fghj)}_{\mu\nu\rho\sigma}(p,q)&=f(p_\mu+q_\mu)\left\{2\left[\bar{g}(p_\nu)\bar{h}(p_\rho)\bar{j}(p_\sigma)+\bar{g}(q_\nu)\bar{h}(q_\rho)\bar{j}(q_\sigma)\right]\right.\nonumber\\
&\hspace{2.5cm}\left.+[\bar{g}(p_\nu)+\bar{g}(q_\nu)][\bar{h}(p_\rho)+\bar{h}(q_\rho)][\bar{j}(p_\sigma)+\bar{j}(q_\sigma)]\right\}
\end{align}
\begin{align}
L^{(fg)}_{\mu\nu}(p,q,k)&=f(p_\mu+q_\mu+k_{\mu})g(2p_\nu+k_{\nu})\\
L^{(fgh)}_{\mu\nu\rho}(p,q,k)&=f(p_\mu+q_\mu+k_{\mu})g(2p_\nu+k_{\nu})
\left[\bar{h}(p_\rho)+\bar{h}(p_\rho+k_{\rho})+\bar{h}(q_\rho)\right]\\
L^{(fghj)}_{\mu\nu\rho\sigma}(p,q,k)&=f(p_\mu+q_\mu+k_{\mu})g(2p_\nu+k_{\nu})\nonumber\\
&\times\left\{\bar{h}(p_\rho)\bar{j}(p_\sigma)+\bar{h}(p_\rho+k_\rho)\bar{j}(p_\sigma+k_\sigma)+\bar{h}(q_\rho)\bar{j}(q_\sigma)\right.\nonumber\\
&\ \ \left.+[\bar{h}(p_\rho)+\bar{h}(p_\rho+k_\rho)+\bar{h}(q_\rho)][\bar{j}(p_\sigma)+\bar{j}(p_\sigma+k_\sigma)+\bar{j}(q_\sigma)]\right\}\hspace{2.cm}
\end{align}
\begin{align}
M^{(fgh)}_{\mu\nu\rho}(p,q,k_1,k_2)&=f(p_\mu+q_\mu+k_{1\mu}+k_{2\mu})g(2p_\nu+k_{1\nu}+2k_{2\nu})h(2p_\rho+k_{2\rho})\\
M^{(fghj)}_{\mu\nu\rho\sigma}(p,q,k_1,k_2)&=f(p_\mu+q_\mu+k_{1\mu}+k_{2\mu})g(2p_\nu+k_{1\nu}+2k_{2\nu})h(2p_\rho+k_{2\rho})\nonumber\\
&\times\left\{\bar{j}(p_\sigma)+\bar{j}(p_\sigma+k_{2\sigma})+\bar{j}(p_\sigma+k_{1\sigma}+k_{2\sigma})+\bar{j}(q_\sigma)\right\}\hspace{2cm}
\end{align}
with $f,g,h,j\in \{s,c\}$.\\
\ \\
\textbf{The $q\bar{q}g$-vertex:}
\begin{align}
V_{1\,\mu}^a(p,q)=-g_0 T^a&\left[\vphantom{\sum\limits_{\nu\neq\mu} }   \right.
\lambda_1\si{p_\mu+q_\mu}+ 2 i \rho_1 \co{p_\mu+q_\mu}\gamma_\mu\nonumber\\
+\sum\limits_{\substack{\nu=1 \\ \nu\neq\mu}}^4 &\left\{ \lambda_2 K^{(sc)}_{\mu\nu}(p,q)      
+2i\rho_2\left(K^{(cc)}_{\mu\nu}(p,q)\gamma_\mu-K^{(ss)}_{\mu\nu}(p,q)\gamma_\nu\vphantom{\frac12}\right)\right\}\nonumber\\
+\frac 13 \sum\limits_{\substack{\nu,\rho=1 \\ \neq(\nu,\rho;\mu)}}^4&  \left\{
 \lambda_3 K^{(scc)}_{\mu\nu\rho}(p,q) 
 +2i\rho_3 \left(K^{(ccc)}_{\mu\nu\rho}(p,q)\gamma_\mu-2K^{(ssc)}_{\mu\nu\rho}(p,q)\gamma_\nu
 \vphantom{\frac12}\right)\right\}\nonumber\\
+\frac 1 9 \sum\limits_{\substack{\nu,\rho,\sigma=1 \\ \neq(\nu,\rho,\sigma;\mu)}}^4 &  \left\{ \lambda_4 K^{(sccc)}_{\mu\nu\rho\sigma}(p,q)      
+2 i\rho_4 \left(K^{(cccc)}_{\mu\nu\rho\sigma}(p,q)\gamma_\mu-3K^{(sscc)}_{\mu\nu\rho\sigma}(p,q)\gamma_\nu\vphantom{\frac12}\right)\right\}
\left.\vphantom{\sum\limits_{\nu\neq\mu} }  \right]
\end{align}
%
%
\textbf{The $q\bar{q}gg$-vertex:}
\begin{align}
V_{2\,\mu\nu}^{ab}&(p,q,k_1,k_2)=ag_0^2 T^aT^b\left\{\vphantom{\frac 12}   \right.
-\frac 12 \lambda_1\delta_{\mu\nu}\co{p_\mu+q_\mu}+  i \rho_1 \delta_{\mu\nu}\si{p_\mu+q_\mu}\gamma_\mu\nonumber\\
+\lambda_2&\left(\vphantom{\frac 12}\right.(1-\delta_{\mu\nu})L^{(ss)}_{\mu\nu}(p,q,k_2)- \frac 12\delta_{\mu\nu}\sum\limits_{\substack{\alpha=1 \\ \alpha\neq\mu}}^4K^{(cc)}_{\mu\alpha}(p,q)\left.\vphantom{\frac 12}\right)\nonumber\\
+\lambda_3&\left(\vphantom{\frac 12 }\right.
\frac 23 (1-\delta_{\mu\nu})
\sum\limits_{\substack{\rho=1 \\ \neq(\rho,\mu,\nu)}}^4 
L^{(ssc)}_{\mu\nu\rho}(p,q,k_2)
-\frac 16 \delta_{\mu\nu}\sum\limits_{\substack{\alpha,\rho=1 \\ \neq(\alpha,\rho,\mu)}}^4K^{(ccc)}_{\mu\alpha\rho}(p,q)
\left.\vphantom{\frac 12}\right)\nonumber\\
+\lambda_4&\left(\vphantom{\frac 11 }\right.
\frac 16(1-\delta_{\mu\nu})
\sum\limits_{\substack{\rho,\sigma=1 \\ \neq(\rho,\sigma,\mu,\nu)}}^4 
L^{(sscc)}_{\mu\nu\rho\sigma}(p,q,k_2)
-\frac 1{18} \delta_{\mu\nu}
\sum\limits_{\substack{\alpha,\rho,\sigma=1 \\ \neq(\alpha,\rho,\sigma,\mu)}}^4
K^{(cccc)}_{\mu\alpha\rho\sigma}(p,q)
\left.\vphantom{\frac 12}\right)\nonumber
\end{align}
\begin{align}
+i\rho_2&\left(\vphantom{\frac 12}
2(1-\delta_{\mu\nu})
\left[L^{(cs)}_{\mu\nu}(p,q,k_2)\gamma_\mu+L^{(sc)}_{\mu\nu}(p,q,k_2)\gamma_\nu\right]\right.\nonumber\\
&\hspace{1cm}+\delta_{\mu\nu}\sum\limits_{\substack{\alpha=1 \\ \alpha\neq\mu}}^4
\left.\left[K^{(sc)}_{\mu\alpha}(p,q)\gamma_\mu+K^{(cs)}_{\mu\alpha}(p,q)\gamma_\alpha\right]\vphantom{\frac 12}\right)
\nonumber\\
+i\rho_3&\left(\vphantom{\frac 12}\right.
\frac 43 (1-\delta_{\mu\nu})
\sum\limits_{\substack{\rho=1 \\ \neq(\rho;\mu,\nu)}}^4 
\left[L^{(csc)}_{\mu\nu\rho}(p,q,k_2)\gamma_\mu+L^{(scc)}_{\mu\nu\rho}(p,q,k_2)\gamma_\nu-L^{(sss)}_{\mu\nu\rho}(p,q,k_2)\gamma_\rho
\right]\nonumber\\
&\hspace{1cm}+\frac 13  \delta_{\mu\nu}
\sum\limits_{\substack{\alpha,\rho=1 \\ \neq(\alpha,\rho;\mu)}}^4
\left.\left[K^{(scc)}_{\mu\alpha\rho}(p,q)\gamma_\mu+2K^{(ccs)}_{\mu\alpha\rho}(p,q)\gamma_\rho
\right]\vphantom{\frac 12}\right)
\nonumber\\
+i\rho_4&\left(\vphantom{\frac 12}\right.
\frac 13 (1-\delta_{\mu\nu})
\sum\limits_{\substack{\rho,\sigma=1 \\ \neq(\rho,\sigma,\mu,\nu)}}^4 
\left[L^{(cscc)}_{\mu\nu\rho\sigma}(p,q,k_2)\gamma_\mu+L^{(sccc)}_{\mu\nu\rho\sigma}(p,q,k_2)\gamma_\nu-2L^{(sssc)}_{\mu\nu\rho\sigma}(p,q,k_2)\gamma_\rho
\right]\nonumber\\
&\hspace{1cm}+\frac 19  \delta_{\mu\nu}
\sum\limits_{\substack{\alpha,\rho,\sigma=1 \\ \neq(\alpha,\rho,\sigma,\mu)}}^4
\left[K^{(sccc)}_{\mu\alpha\rho\sigma}(p,q)\gamma_\mu+3K^{(ccsc)}_{\mu\alpha\rho\sigma}(p,q)\gamma_\rho
\right]\left.\left.\vphantom{\frac 12}\right)\right\}
\end{align}
%
%
\textbf{The $q\bar{q}ggg$-vertex:}
\begin{align}
V_{3\,\mu\nu\rho}^{abc}&(p,q,k_1,k_2,k_3)=a^2g_0^3 T^aT^bT^c
\left\{\vphantom{\frac 12}   \right.
\frac 16\delta_{\mu\nu}\delta_{\mu\rho}\left( \lambda_1\si{p_\mu+q_\mu}+ 2 i \rho_1 \co{p_\mu+q_\mu}\gamma_\mu\right)\nonumber\\
+\lambda_2&\left(\vphantom{\frac 12}\right.
\frac 16 \delta_{\mu\nu}\delta_{\mu\rho}
\sum\limits_{\substack{\alpha=1 \\ \alpha\neq\mu}}^4
K^{(sc)}_{\mu\alpha}(p,q)\nonumber\\
&+\frac 12 \delta_{\mu\nu}(1-\delta_{\mu\rho})
L^{(cs)}_{\mu\rho}(p,q,k_3)
+\frac 12 \delta_{\nu\rho}(1-\delta_{\mu\nu})
L^{(sc)}_{\mu\nu}(p,q,k_2+k_3)
\left.\vphantom{\frac 12}\right)\nonumber\\
+\lambda_3&\left(\vphantom{\frac 12 }\right.
-\frac 23 (1-\delta_{\mu\nu})(1-\delta_{\mu\rho})(1-\delta_{\nu\rho})
M^{(sss)}_{\mu\nu\rho}(p,q,k_2,k_3)
+\frac 1{18} \delta_{\mu\nu}\delta_{\mu\rho}
\sum\limits_{\substack{\alpha,\beta=1 \\ \neq(\alpha,\beta,\mu)}}^4K^{(scc)}_{\mu\alpha\beta}(p,q)\nonumber\\
&+\frac 13\delta_{\mu\nu}(1-\delta_{\mu\rho})
\sum\limits_{\substack{\alpha=1 \\ \neq(\alpha,\mu,\rho)}}^4
L^{(csc)}_{\mu\rho\alpha}(p,q,k_3)
+\frac 13 \delta_{\nu\rho}(1-\delta_{\mu\nu})
\sum\limits_{\substack{\alpha=1 \\ \neq(\alpha,\mu,\nu)}}^4
L^{(scc)}_{\mu\nu\alpha}(p,q,k_2+k_3)
\left.\vphantom{\frac 12}\right)\nonumber\\
+\lambda_4&\left(\vphantom{\frac 12 }\right.
-\frac 13 (1-\delta_{\mu\nu})(1-\delta_{\mu\rho})(1-\delta_{\nu\rho})
\sum\limits_{\substack{\sigma=1 \\ \neq(\sigma,\mu,\nu,\rho)}}^4
M^{(sssc)}_{\mu\nu\rho\sigma}(p,q,k_2,k_3)\nonumber\\
&+\frac {1}{54} \delta_{\mu\nu}\delta_{\mu\rho}
\sum\limits_{\substack{\alpha,\beta,\sigma=1 \\ \neq(\alpha,\beta,\sigma,\mu)}}^4
K^{(sccc)}_{\mu\alpha\beta\sigma}(p,q)\nonumber\\
&+\frac 1{12}\delta_{\mu\nu}(1-\delta_{\nu\rho})
\sum\limits_{\substack{\alpha,\sigma=1 \\ \neq(\alpha,\sigma;\mu,\rho)}}^4
L^{(cscc)}_{\mu\rho\alpha\sigma}(p,q,k_3)
+\frac{1}{12}\delta_{\nu\rho}(1-\delta_{\mu\nu})
\sum\limits_{\substack{\alpha,\sigma=1 \\ \neq(\alpha,\sigma;\mu,\nu)}}^4
L^{(sccc)}_{\mu\nu\alpha\sigma}(p,q,k_2+k_3)
\left.\vphantom{\frac 12}\right)\nonumber
\end{align}
%
%
\begin{align}
+i\rho_2&\left(\vphantom{\frac 12}\right.
\frac 13 \delta_{\mu\nu}\delta_{\mu\rho}
\sum\limits_{\substack{\alpha=1 \\ \alpha\neq\mu}}^4
\left[K^{(cc)}_{\mu\alpha}(p,q)\gamma_\mu-K^{(ss)}_{\mu\alpha}(p,q)\gamma_\alpha\right]
\nonumber\\
&+\delta_{\mu\nu}(1-\delta_{\mu\rho})
\left[L^{(cc)}_{\mu\rho}(p,q,k_3)\gamma_\rho-L^{(ss)}_{\mu\rho}(p,q,k_3)\gamma_\mu\right]\nonumber\\
&+\delta_{\nu\rho}(1-\delta_{\mu\nu})
\left[L^{(cc)}_{\mu\rho}(p,q,k_2+k_3)\gamma_\mu-L^{(ss)}_{\mu\rho}(p,q,k_2+k_3)\gamma_\nu\right]
\left.\vphantom{\frac 12}\right)\nonumber\\
+i\rho_3&\left(\vphantom{\frac 12}\right.
-\frac 43 (1-\delta_{\mu\nu})(1-\delta_{\mu\rho})(1-\delta_{\nu\rho})\nonumber\\
&\times\left[\vphantom{K^{()}_{\beta}}M^{(css)}_{\mu\nu\rho}(p,q,k_2,k_3)\gamma_\mu+
M^{(scs)}_{\mu\nu\rho}(p,q,k_2,k_3)\gamma_\nu+
M^{(ssc)}_{\mu\nu\rho}(p,q,k_2,k_3)\gamma_\rho
\right]\nonumber\\
&+\frac 19 \delta_{\mu\nu}\delta_{\mu\rho}
\sum\limits_{\substack{\alpha,\beta=1 \\ \neq(\alpha,\beta,\mu)}}^4 
\left[K^{(ccc)}_{\mu\alpha\beta}(p,q)\gamma_\mu 
-K^{(ssc)}_{\mu\alpha\beta}(p,q)\gamma_\alpha
-K^{(scs)}_{\mu\alpha\beta}(p,q)\gamma_\beta\right]
\nonumber\\
&-\frac 23 \delta_{\mu\nu}(1-\delta_{\mu\rho})
\sum\limits_{\substack{\alpha=1 \\ \neq(\alpha,\mu,\rho)}}^4 
\left[L^{(ssc)}_{\mu\rho\alpha}(p,q,k_3)\gamma_\mu
-L^{(ccc)}_{\mu\rho\alpha}(p,q,k_3)\gamma_\rho
+L^{(css)}_{\mu\rho\alpha}(p,q,k_3)\gamma_\alpha\right]
\nonumber\\
&-\frac 23 \delta_{\nu\rho}(1-\delta_{\mu\nu})
\sum\limits_{\substack{\alpha=1 \\ \neq(\alpha,\mu,\nu)}}^4 
\left[-L^{(ccc)}_{\mu\nu\alpha}(p,q,k_2+k_3)\gamma_\mu
+L^{(ssc)}_{\mu\nu\alpha}(p,q,k_2+k_3)\gamma_\nu
\right.\nonumber\\
&\hspace{4cm}\left.
+L^{(scs)}_{\mu\nu\alpha}(p,q,k_2+k_3)\gamma_\alpha\right]
\left.\vphantom{\frac 12}\right)\nonumber\\
+i\rho_4&\left(\vphantom{\frac 12}\right.
-\frac 2 3 (1-\delta_{\mu\nu})(1-\delta_{\mu\rho})(1-\delta_{\nu\rho})
\!\!\!\!\!\!\sum\limits_{\substack{\sigma=1 \\ \neq(\sigma,\mu,\nu,\rho)}}^4 
\!\!\!\!\!\!\left[M^{(cssc)}_{\mu\nu\rho\sigma}(p,q,k_2,k_3)\gamma_\mu+
M^{(scsc)}_{\mu\nu\rho\sigma}(p,q,k_2,k_3)\gamma_\nu
\vphantom{K^{()}_\beta}\right.\nonumber\\
&\hspace{5.6cm}\left.
+M^{(sscc)}_{\mu\nu\rho\sigma}(p,q,k_2,k_3)\gamma_\rho
-M^{(ssss)}_{\mu\nu\rho\sigma}(p,q,k_2,k_3)\gamma_\sigma
\vphantom{K^{()}_\beta}\right]\nonumber\\
&+\frac 1{27} \delta_{\mu\nu}\delta_{\mu\rho}
\sum\limits_{\substack{\alpha,\beta,\sigma=1 \\ \neq(\alpha,\beta,\sigma,\mu)}}^4 
\left[K^{(cccc)}_{\mu\alpha\beta\sigma}(p,q)\gamma_\mu 
-K^{(sscc)}_{\mu\alpha\beta\sigma}(p,q)\gamma_\alpha
\right.\nonumber\\
&\hspace{3.3cm}\left.
-K^{(scsc)}_{\mu\alpha\beta\sigma}(p,q)\gamma_\beta
-K^{(sccs)}_{\mu\alpha\beta\sigma}(p,q)\gamma_\sigma\right]
\nonumber\\
&-\frac 1 6\delta_{\mu\nu}(1-\delta_{\mu\rho})
\sum\limits_{\substack{\alpha,\sigma=1 \\ \neq(\alpha,\sigma,\mu,\rho)}}^4 
\left[L^{(sscc)}_{\mu\rho\alpha\sigma}(p,q,k_3)\gamma_\mu
-L^{(cccc)}_{\mu\rho\alpha\sigma}(p,q,k_3)\gamma_\rho
\right.\nonumber\\
&\hspace{4.5cm}\left.
+L^{(cssc)}_{\mu\rho\alpha\sigma}(p,q,k_3)\gamma_\alpha
+L^{(cscs)}_{\mu\rho\alpha\sigma}(p,q,k_3)\gamma_\sigma\right]
\vphantom{\frac 12}\nonumber\\
&+\frac 1 6\delta_{\nu\rho}(1-\delta_{\mu\nu})
\sum\limits_{\substack{\alpha,\sigma=1 \\ \neq(\alpha,\sigma,\mu,\nu)}}^4 
\left[L^{(cccc)}_{\mu\nu\alpha\sigma}(p,q,k_2+k_3)\gamma_\mu
-L^{(sscc)}_{\mu\nu\alpha\sigma}(p,q,k_2+k_3)\gamma_\nu
\right.\nonumber\\
&\hspace{4.5cm}\left.
+L^{(scsc)}_{\mu\nu\alpha\sigma}(p,q,k_2+k_3)\gamma_\alpha
+L^{(sccs)}_{\mu\nu\alpha\sigma}(p,q,k_2+k_3)\gamma_\sigma\right]
\left.\left.\vphantom{\frac 12}\right)\right\}
\end{align}
\newpage
Purely gluonic Feynman rules (gluon propagator and $ggg$-vertex) as well as
contributions to the vertex Feynman rules coming from the clover term containing the improvement coefficient $\csw$ are unaffected by the fermion formulations and thus identical to the Wilson case (see Ref.~\cite{Aoki:2003sj}).
\section{Perturbative Determination of $\csw$}
\begin{figure}
\centering
\begin{tabular}{ c c  c}
(a) &(b) & (c)\\
\begin{tikzpicture}
  \begin{feynman}
  \vertex[draw,circle,fill,inner sep=1.pt] (a){};
  \vertex [right= 2.2cm of a, inner sep=2pt] (b){};
  \vertex [above left = 1.5cm and 1.5cm of a] (i1);
  \vertex [below left = 1.5cm and 1.5cm of a] (i2);
  \draw[draw=none] (i2) to node[pos=0.5,draw,circle,fill,inner sep=1.] (h2){} (a);
  \draw[draw=none] (i1) to node[pos=0.5,draw,circle,fill,inner sep=1.] (h1){} (a);
  \diagram*{
   (i2) -- [fermion,edge label'=$p$] (h2) --[fermion,edge label'=$p+k$] (a),
   (a) -- [fermion,edge label'=$q+k$] (h1) -- [fermion,edge label'=$q$] (i1),
   (a) -- [gluon,rmomentum={[arrow shorten=0.2,xshift=0.3cm]$q-p$}]  (b),
   (h1) --[gluon,momentum'=$k$] (h2),
   };
  \end{feynman}
\end{tikzpicture}

&
\begin{tikzpicture}
  \begin{feynman}
  \vertex[draw,circle,fill,inner sep=1.pt] (a){};
  \vertex [right= 2.2cm of a, inner sep=2pt] (b){};
  \vertex [above left = 1.5cm and 1.5cm of a] (i1);
  \vertex [below left = 1.5cm and 1.5cm of a] (i2);
  \draw[draw=none] (i2) to node[pos=0.5,draw,circle,fill,inner sep=1.] (h2){} (a);
  \draw[draw=none] (i1) to node[pos=0.5,draw,circle,fill,inner sep=1.] (h1){} (a);
  \diagram*{
   (i2) -- [fermion,edge label'=$p$] (h2) --[gluon,momentum'=$p-q-k$] (a),
   (a) -- [gluon,rmomentum'=$k$] (h1) -- [fermion,edge label'=$q$] (i1),
   (a) -- [gluon,rmomentum={[arrow shorten=0.2,xshift=0.3cm]$q-p$}] (b),
   (h2) --[fermion,edge label=$q+k $] (h1),
   };
  \end{feynman}
\end{tikzpicture}
&

\begin{tikzpicture}
  \begin{feynman}
  \vertex[draw,circle,fill,inner sep=1.pt] (a){};
  \vertex [right= 2.2cm of a, inner sep=2pt] (b){};
  \vertex [above left = 1.5cm and 1.5cm of a] (i1);
  \vertex [below left = 1.5cm and 1.5cm of a] (i2);
  \draw[draw=none] (a) to node[pos=0.5,draw,circle,fill,inner sep=1.] (h){} (b);
  \diagram*{
   (i2) -- [fermion,edge label'=$p$, near start]  (a),
   (a) -- [fermion,edge label'=$q$, near end] (i1),
   (h) -- [gluon,rmomentum={[arrow shorten=0.,xshift=0.3cm]$q-p$}] (b),
    a--[gluon,half right,momentum'={[arrow shorten=0.3]$p-q+k$}] (h)--[gluon, half right,momentum'={[arrow shorten=0.3]$k$}] a
   };
  \end{feynman}
\end{tikzpicture}
\\

(d) & (e) & (f) \\
\begin{tikzpicture}
  \begin{feynman}
  \vertex[draw,circle,fill,inner sep=1.pt] (a){};
  \vertex [right= 2.cm of a, right,inner sep=1.5pt] (b){};
  \vertex [above right= 1.3cm and .85cm of a, inner sep=0.cm,right] (c){};
  \vertex [below right= 1.3cm and .85cm of a, inner sep=0.cm, right] (d){};
  \vertex [above left = 1.5cm and 1.5cm of a] (i1);
  \vertex [below left = 1.5cm and 1.5cm of a] (i2);
  \diagram*{
   (i2) --[fermion,edge label'=$p$, near start] (a),
   (a) -- [fermion,edge label'=$q$, near end] (i1),
   (a) -- [gluon,rmomentum={[arrow shorten=0.2,xshift=0.3cm]$q-p$}] (b),
   a --[gluon,out=-105,in=-37, loop, min distance=1.4cm, momentum'={[arrow shorten=0.3]$k$}] a
   };
  \end{feynman}
\end{tikzpicture}

&
\begin{tikzpicture}
  \begin{feynman}
  \vertex[draw,circle,fill,inner sep=1.pt] (a){};
  \vertex [right= 2.2cm of a, inner sep=2pt] (b){};
  \vertex [above left = 1.5cm and 1.5cm of a] (i1);
  \vertex [below left = 1.5cm and 1.5cm of a] (i2);
  \draw[draw=none] (i1) to node[pos=0.5,draw,circle,fill,inner sep=1.] (h1){} (a);
  \diagram*{
   (i2)--[fermion,edge label'=$p$, near start] (a),
   (a) -- [fermion,edge label=$q+k$] (h1) -- [fermion,edge label'=$q$] (i1),
   (a) -- [gluon,rmomentum={[arrow shorten=0.2,xshift=0.3cm]$q-p$}] (b),
   (a) --[gluon, out=60, in =30, distance =0.6cm, rmomentum'={[arrow shorten=0.3]$k$}] (h1),
   };
  \end{feynman}
\end{tikzpicture}
&
\begin{tikzpicture}
  \begin{feynman}
  \vertex[draw,circle,fill,inner sep=1.pt] (a){};
  \vertex [right= 2.2cm of a, inner sep=2pt] (b){};
  \vertex [above left = 1.5cm and 1.5cm of a] (i1);
  \vertex [below left = 1.5cm and 1.5cm of a] (i2);
  \draw[draw=none] (i2) to node[pos=0.5,draw,circle,fill,inner sep=1.] (h2){} (a);
  \diagram*{
   (i2) -- [fermion,edge label'=$p$] (h2)--[fermion, edge label=$p-k$] (a),
   (a) -- [fermion,edge label'=$q$, near end] (i1),
   (a) -- [gluon,rmomentum={[arrow shorten=0.2,xshift=0.3cm]$q-p$}] (b),
   (h2) --[gluon, out=-30, in =-60, distance =0.6cm, momentum'={[arrow shorten=0.3]$k$}] (a),
   };
  \end{feynman}
\end{tikzpicture}
\\
\end{tabular}
\caption{The six one-loop diagrams contributing to the vertex function in lattice perturbation theory. }
\label{figdiagrams}
\end{figure}
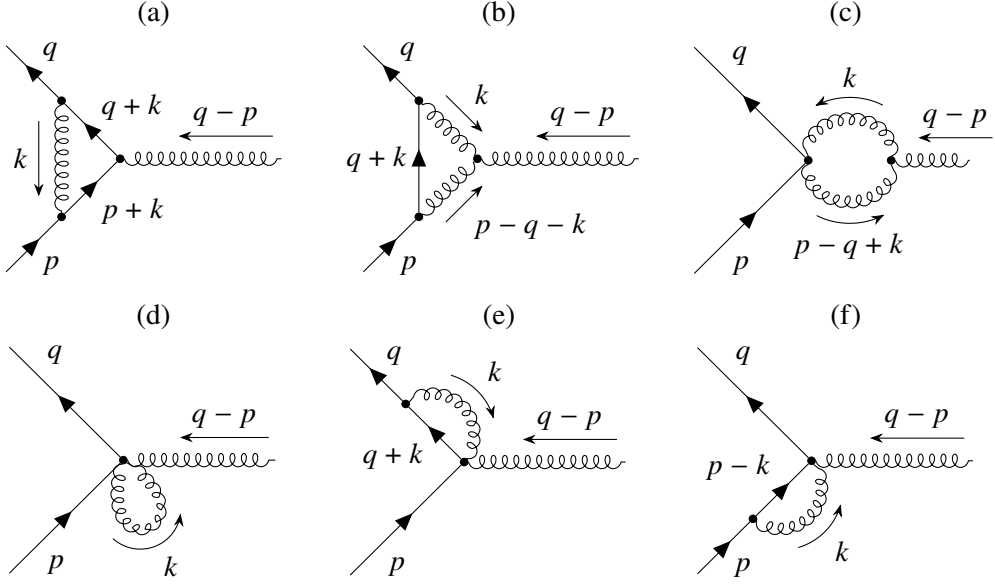

Adding the usual clover-term to the  Brillouin action we obtain the $\mathcal{O}(a)$-improved Brillouin action (with chromo-hermitean field strength $F_{\mu \nu}$ for fixed $\mu<\nu$):
\begin{align}
\mathcal{S}_\text{Brillouin}^\text{Clover}[\overline{\psi},\psi]=\mathcal{S}_\text{Brillouin}[\overline{\psi},\psi]+c_\text{SW}\cdot \sum\limits_x\sum\limits_{\mu<\nu}\bar{\psi}(x)\frac 12 \sigma_{\mu\nu}F_{\mu\nu}(x)\psi(x).
\end{align}
The improvement coefficient has a perturbative expansion $\csw=\csw^{(0)}+g_0^2\csw^{(1)}+\mathcal{O}(g_0^4)$ and at tree level in both the Wilson and the Brillouin case:
\begin{align}
\csw^{(0)}=1(=r)
\end{align}
At one-loop level $\csw^{(1)}$ is calculated from the one-loop vertex function $\Lambda_\mu^{a(1)}$ comprised of six one-loop Feynman diagrams (see Figure \ref{figdiagrams}).
The sum of all diagrams results in a finite integral, while diagrams (a),(b),(c),(e), and (f) on their own lead to IR-divergent integrals. 
To see the cancellation  of the divergencies explicitly, we used a small gluon mass $\mu$ as a regulator (following Ref.~\cite{Aoki:2003sj}) and split off an analytically solvable divergent integral from the diagrams.
Table \ref{tab} shows the results for each diagram for $N_c=3$ and $r=1$, comparing the Wilson and the Brillouin actions.
\begin{table}[!h!]
\centering
\def\arraystretch{1.}
\begin{tabular}{|c|c|c|c|}
\hline
Diagram & Divergent part & Constant part Wilson & Constant part Brillouin\\
\hline
(a) & $-L/3$ & $0.004569626(1)$  & $0.0047576939 (1)$ \\
(b) & $-9L/2$ & $0.083078349 (1)$  & $0.055554134 (1)$ \\
(c) & $+9L/2$ & $-0.081307544 (1)$  & $-0.057741323 (1)$ \\
(d) & $0$ & $0.297394534 (1)$ & $0.142461144 (1)$ \\
(e) & $L/6$ & $- 0.017573359 (1) $& $-0.010702925 (1)$ \\
(f) & $L/6$ & $- 0.017573359 (1)$ & $-0.010702925 (1)$\\
\hline
Sum & $0$ & $0.26858825 (1)$  & $0.12362580 (1)$\\
\hline
\end{tabular}
\caption{Divergent and constant contributions from each diagram for $N_c=3$ and $r=1$.
The logarithmic divergence is encoded in $L:=\frac{1}{16\pi^2}\ln\left(\tfrac{\pi^2}{\mu^2}\right)$.}
\label{tab}
 \end{table}
\ \\
\newpage
\noindent 
The sum of all diagrams finally results in the following value
\begin{align}
{\csw}^{(1)}_\text{Brillouin} = 0.045785517(3) N_c -0.041192255(3) \dfrac{1}{N_c} =0.12362580(1),
\end{align}
where we have set $N_c=3$ in the last step.
Compare to 
\begin{align}
{\csw}^{(1)}_\text{Wilson} =0.0988424712 (4) N_c - 0.083817496 (3) \dfrac{1}{N_c}  = 0.26858825(1)
\end{align}
 from Ref.~\cite{Aoki:2003sj}, one sees that ${\csw}^{(1)}_\text{Brillouin}$ is about
half of ${\csw}^{(1)}_\text{Wilson}$ for any value of $N_c$.

\end{document}